# Viscoelastic interfacial structures in undersaturated calcium solutions under nanoconfinement

Shurui Miao*[1], Timothy S. Groves[1], Kieran J. Agg[1], Susan Perkin*[1]

[1] Department of Chemistry, Physical & Theoretical Chemistry Laboratory, University of Oxford, United Kingdom.

Abstract:

Calcium mineralisation underpins processes ranging from biomineralisation to scalable carbon storage, yet the earliest stages of nucleation remain unresolved. Here we investigate calcium-containing solutions under nanoscopic confinement while remaining undersaturated with respect to bulk calcium carbonate precipitation. Using sensitive surface force measurements, we identify a long-range non-DLVO repulsive interaction specific to calcium solutions, which we attribute to hydrated ion networks at mineral interfaces. These interfacial networks exhibit viscoelastic behaviour distinct from the bulk solution, persist across a broad pH range, and can be disrupted by competing ions with strong surface affinity. Our findings provide experimental evidence that stable hydrated ion networks can emerge at mineral interfaces under undersaturated conditions, supporting the possibility that interfaces and confinement stabilise intermediate ion assemblies prior to nucleation. These results bridge classical and nonclassical descriptions of mineral nucleation and highlight the central role of interfaces and confinement in directing crystallisation pathways.

Nucleation of calcium-based minerals is of profound scientific, medical, and industrial importance. It governs the formation of geological deposits[1] and biominerals like shells and bone,[2] as well as underpinning emerging carbon storage technologies.[3, 4] Therefore, the study of calcium minerals has a long and extensive history, especially around understanding their rich polymorphism and the intricate physicochemical conditions under which mineralisation occurs. This sustained effort has established calcium minerals as canonical model systems for gaining fundamental insights into crystallisation mechanisms and phase stability.

Decades of research have sought to determine whether crystal formation follows the classical nucleation theory (CNT).[5-14] Although it has successfully described many crystallisation processes, it often fails in complex systems, including cases where amorphous intermediates form at concentrations significantly below classical predictions.[9, 15] Such observations have motivated alternative models in which the early stages of crystallisation are governed by formation of locally-stable structures, introducing a free energy minimum rather than only a maximum (as in CNT) along the assembly coordinate. Part of the nucleation driving force is released during formation of the metastable intermediates, which may be long lived and enable early stages of structural ordering corresponding to germination of specific crystalline polymorphs.[9]

Calcium carbonate ($CaCO_3$) mineralisation is of particular complexity and significance.[5-12, 16] Owing to its extremely low solubility in water ($K_{sp} \sim 10^{-9}$), any study of calcium-containing aqueous solutions in equilibrium with the atmosphere must consider their thermodynamic stability with respect to $CaCO_3$ precipitation. The concentration of dissolved atmospheric $CO_2$ in pure water is on the order of $10^{-5}$ M. Although the speciation of dissolved inorganic carbon is strongly pH-dependent, small but finite concentrations of bicarbonate and carbonate ions are always present ($\sim 10^{-10}$ M carbonate at pH = 6, full calculation in SI).[17] Therefore, calcium-containing solutions in equilibrium with air contain all of the chemical constituents required for $CaCO_3$ precipitation, while enabling precise control over the degree of saturation through variables including the pH and calcium concentration.

In nature, crystallisation often occurs at the solid-liquid interface and under confinement, where physicochemical properties can differ significantly from the bulk.[1, 18-20] Within confined spaces, the presence of solid-solution interfaces and overlapping electrical double layers can strongly influence molecular transport,[21] local solute concentrations and speciation,[22] local electrical potentials,[23] and even the structuring of the surrounding solution.[24] As a result, confinement introduces an additional dimension of complexity to crystallisation mechanisms and is central to our understanding of solution processes in both natural and engineered systems. Despite recent advancement in experimental capabilities such as liquid-phase electron microscopy and in situ x-ray scattering,[5, 10] it remains a challenge to study heterogeneous nucleation and the effect of confinement.

The Surface Force Balance (SFB, also known as the Surface Force Apparatus) is a versatile experimental method to study the effect of confinement on solution behaviours (Figure 1A).[25] The ability of the SFB to produce carefully controlled confinement conditions makes it an ideal tool to study nucleation from calcium solutions. Recently, several studies have made use of the SFB to investigate and control the growth of confined calcium minerals,[26-32] however the undersaturated regime has largely escaped this attention. One work from Israelachvili and Pashley has studied $CaCl_2$ solutions at concentrations up to 5 M in equilibrium with air.[33] They report repulsive forces across the electrolyte extending to several nanometres for highly concentrated solutions (>3 M). Here, we consider the effect of confinement on $Ca^{2+}$-containing aqueous electrolytes equilibrated with atmospheric $CO_2$ at concentrations below bulk saturation with respect to $CaCO_3$ (full calculation in SI). We find that the calcium-containing aqueous solution forms a viscoelastic film when confined to films below $\sim$ 50 nm. The observation is specific to $Ca^{2+}$ solutions and is subtly dependent on concentration, pH and co-ions. We propose that the negative electric potential from the mineral surfaces concentrates the cations in the near-surface region and leads to formation of a calcium- and carbonate-containing hydrated network. The interfacial film may relate to previous reports of prenucleation clusters,[14] liquid-liquid phase separations,[5] and amorphous aggregates in the bulk solution.[13] This provides insight into calcium mineral nucleation and may advance our ability to direct mineralisation under confinement.

**Interfacial calcium networks are viscoelastic**

A schematic of the SFB setup is shown in Figure 1A. We chose muscovite mica as the model mineral surface since it can be reliably processed into sufficiently large (*i.e.*, > $cm^2$) and atomically smooth surfaces. This is critical since any surface roughness beyond the nanometre scale will convolute and mask signatures of nanoscopic aggregates.[9, 10, 34-38] Mica becomes negatively charged when immersed in aqueous solutions due to the dissolution of surface potassium ions. As a benchmark, a representative profile of measured mean interaction free energy per unit area ($W$) as a function of mica-mica separation distance ($D$) across ultrapure water equilibrated with air is shown in Figure 1B.

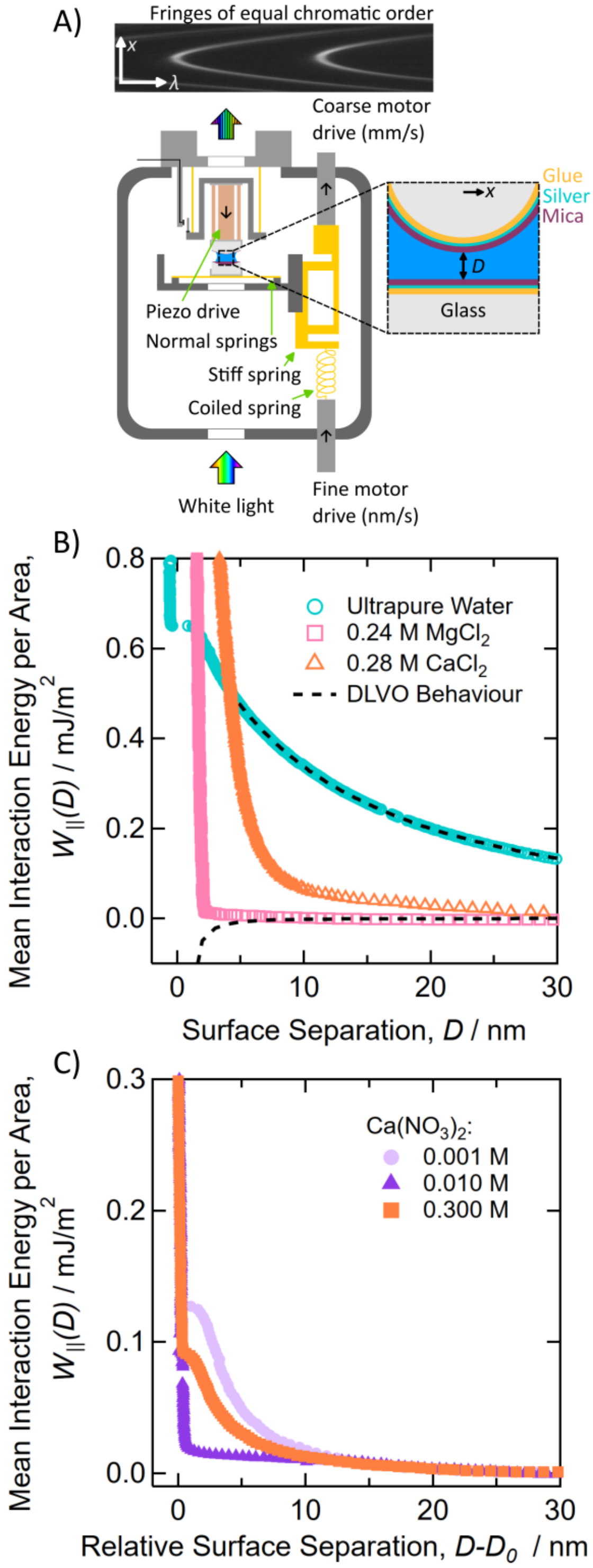

**Figure 1.** A) Schematic of the Surface Force Balance (SFB). Using interferometry and calibrated normal springs, we measure forces between mica sheets across electrolyte solutions as a function of separation. The cross-cylinder geometry allows us to apply the Derjaguin approximation to interpret the measured equilibrium force as interaction energy per unit area between parallel plates.[25] B) Interaction between mica sheets across ultrapure water, 0.24 M $MgCl_2$ and 0.28 M $CaCl_2$ solutions. The predicted DLVO behaviour for water and for the $MgCl_2$ solution are shown as dashed lines. For water, the fitted screening length ($\lambda$) corresponds to $10^{-4}$ M monovalent ions. The $MgCl_2$ solution also exhibits classical behaviour with negligible EDL repulsion. In contrast, a non-DLVO long-range

repulsive interaction is measured across the $CaCl_2$ solution. C) Interaction between mica sheets across a series of $Ca(NO_3)_2$ solutions. The two relatively more dilute solutions are consistent with the DLVO theory (purple). However, at 0.3 M, a non-DLVO long-range repulsive interaction is again observed, like $CaCl_2$ solutions (orange). Curves are plotted in relative surface separation for clarity.

Two distinct features are observed. Firstly, a monotonic exponential repulsive interaction is measured at large separations. For two identical charged surfaces interacting across the electrolyte, a repulsive term arises from the excess osmotic pressure due to overlap of the diffuse layers of counter ions associated with each surface. This is often referred to as the electrical double layer (EDL) force. The charge distribution and thus the repulsive force arising from excess osmotic pressure can be calculated using the Poisson-Boltzmann equation. To account for ion adsorption to the charged surface, we also adopt the charge regulation approximation.[39] Secondly, at small separations of the surfaces ($D \sim 2$ nm), we measure a discontinuity (a jump in $D$) in the profile before the mica surfaces are brought into contact. That is, at a certain separation, the EDL repulsion between the negatively charged mica surfaces is suddenly overcome by a relatively short-ranged attractive interaction. The jump corresponds to a displacement greater than 1 nm, significantly bigger than any molecular features of water. This is consistent with literature and attributed to van der Waals (vdW) attraction between pristine mica surfaces,[40, 41] which we estimated using the Hamaker approach (SI). Overall, the total measured interaction profile can be described by the sum of these two contributions, as described by the Derjaguin–Landau–Verwey–Overbeek (DLVO) theory and expressed as:

$$W_{||} = -\frac{A}{12\pi D^2} + 2\epsilon_0\epsilon\kappa_D\psi_{\text{eff}}^2\frac{e^{-\kappa_D D}}{1+(1-2p)e^{-\kappa_D D}} \qquad \text{Eq. 1}$$

Here, $A$ is the mica-electrolyte-mica Hamaker constant, $\epsilon_0$ is the permittivity of free space, $\epsilon$ is the relative permittivity, $\psi_{\text{eff}}$ is the effective surface potential, $\kappa_D^{-1}$ is the Debye-Hückel screening length, and $p$ is a dimensionless charge regulation parameter ranging from $p = 0$ (constant potential boundary condition) to $p = 1$ (constant charge boundary condition). The Debye-Hückel screening length relates to ion concentration by:

$$\kappa_D = \sqrt{\frac{2N_A e^2 I}{\epsilon_0\epsilon k_B T}} \qquad \text{Eq. 2}$$

where $N_A$ is the Avogadro's constant, $e$ is the electron charge, $I$ is the ionic strength, $k_B$ is the Boltzmann's constant, and $T$ is the temperature.

By fitting Equation 1 to our measured interactions across ultrapure water, we can extract the effective screening length (49.6 ± 0.5 nm). Assuming the ionic strength arises from majority monovalent ions, this corresponds to $\sim 10^{-4}$ M salt concentration. This low background concentration likely arises from leaching of ions from glassware during setup, as well as dissolved $CO_2$ and $K^+$ from the mica surfaces.[17, 42] This confirms that our setup can accurately and reliably detect subtle and long-range repulsive forces.

The Debye-Hückel theory predicts that increasing the ionic strength reduces the screening length, such that the EDL repulsion between mica surfaces becomes shorter ranged and weaker at higher ionic strength. Figure 1B also shows force profiles measured in aqueous solutions of 0.24 M $MgCl_2$ and 0.28 M $CaCl_2$. At these concentrations, the theoretical Debye-Hückel screening length is $\sim 0.3$ nm. The $MgCl_2$ experimental trace clearly follows this prediction at long distances with negligible repulsive interactions measured until the surfaces are very close (~2 nm apart) where a strong repulsive interaction is measured, likely arising from steric repulsions between hydrated ions adsorbed at the mica surfaces.[43] In contrast, a non-DLVO long-range repulsive interaction is measured between mica surfaces across a solution of 0.28 M $CaCl_2$ (Figure 1B). The qualitative difference between $MgCl_2$ and $CaCl_2$ highlights that our observation is cation-specific, and not a general phenomenon for divalent electrolyte solutions at this concentration.

To check for anion-specific effects, we compared $Ca(NO_3)_2$ to the $CaCl_2$ solutions. Figure 1C shows the measured interactions between mica surfaces across $Ca(NO_3)_2$ solutions at a range of concentrations. At the lowest salt concentration (0.001 M), the measured interaction profile is comparable to that of ultrapure water (Figure 1B). We observe both the long-range EDL repulsion and the vdW jump-in at short mica-mica separation. The experimental interaction profile can be fitted by the DLVO theory, and the measured screening length agrees quantitatively with the Debye-Hückel theory (see SI). Similarly, at an intermediate salt concentration (0.01 M), experimental results match the DLVO description, with the range of EDL repulsion now much weaker and shorter in range. In contrast, at a higher salt concentration of 0.3 M, a long-range repulsive force has emerged comparable to our results from $CaCl_2$ in Figure 1B. This indicates the long-range repulsion, apparent in both $CaCl_2$ and $Ca(NO_3)_2$ solutions, appears to be a feature arising from the presence of $Ca^{2+}$ cations. Our results confirm that the non-DLVO long-range force is only detectable above 0.1 M $Ca^{2+}$, below which the interaction between mica surfaces follows the classical DLVO description.

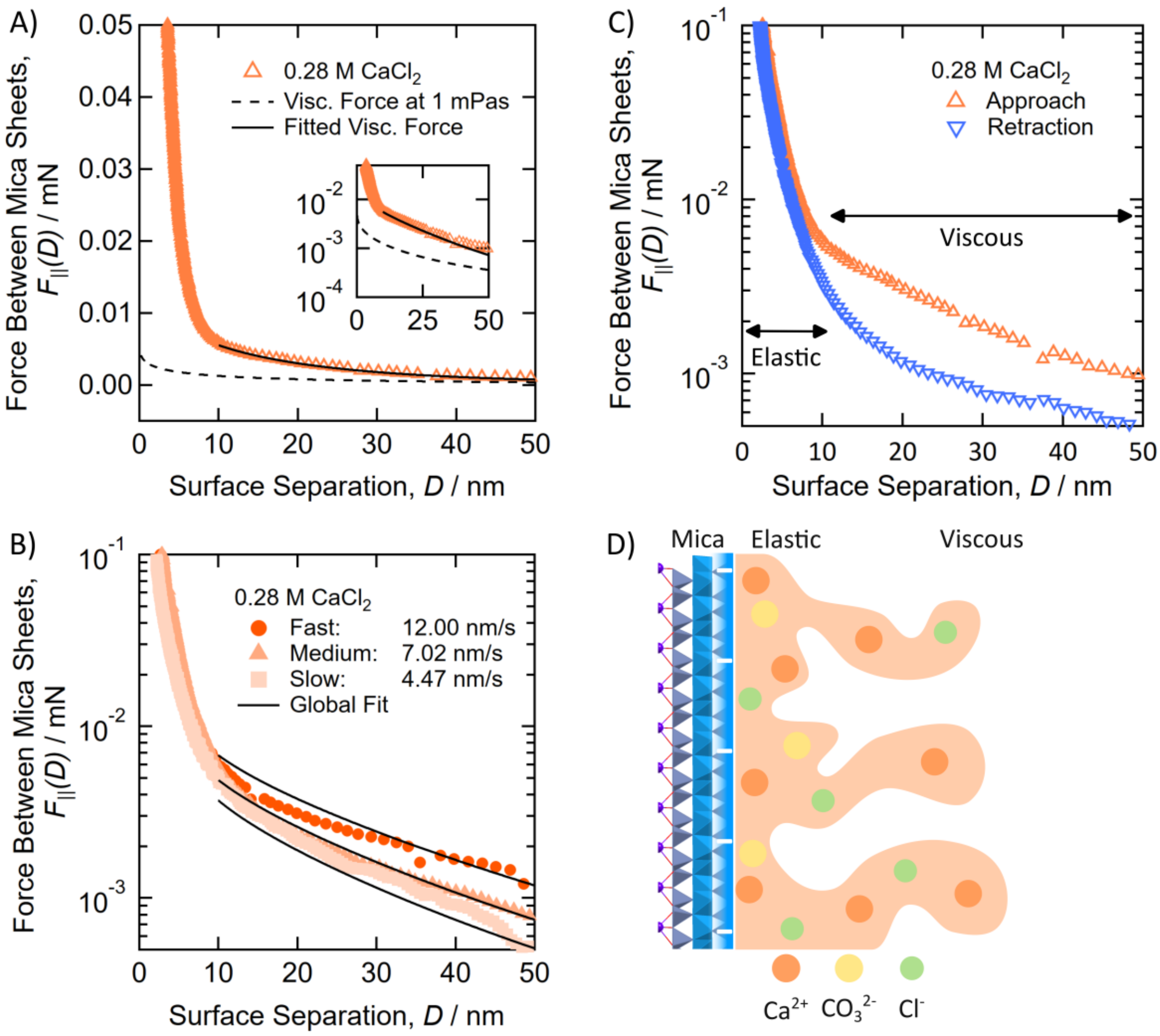

**Figure 2.** A) The non-DLVO repulsive force between mica sheets across the $CaCl_2$ solution is fitted as a viscous force according to Equations 3 and 4. The black dashed line shows the expected force based on bulk viscosity (1.06 mPas), which clearly does not capture the measured force. Our fit accurately captures the long-range force and fails when the separation $D < 10$ nm, where compressibility and other effects dominate the measured force. We fitted the force instead of mean interaction free energy since the viscous force is non-equilibrium in nature. B) Experimental data measured at different approach speeds. Lines show a global fit using Equation 3 and 4 with one set of the parameters $H$, $b$, and $\lambda$. Despite some scatter, the fit is consistent with experimental data and demonstrates that the speed dependence of viscous force is well described by our model. C) Reversibility of the measured force during an approach-retraction cycle over $10^2$ s. For separations $D < 10$ nm, the force is reversible, indicating predominantly elastic behaviour. At larger separations, hysteresis between the approach and retraction curves suggests significant viscous dissipation. D) Schematic of the interfacial hydrated network with elastic and viscous regimes.

To understand the origin of the non-DLVO long-range repulsion, we took inspiration from studies of nucleation mechanisms. In the context of calcium carbonate mineralisation, regardless of the nature of the nucleation pathways, local enrichment of calcium and carbonate is recognised as a necessary intermediate before crystallisation.[5, 9] We hypothesise that the long-range and rate-dependent interaction observed may be attributed to slow dynamics of an interfacial hydrated ion network displaying longer relaxation timescales than simple aqueous solutions. We begin by attempting to model slow relaxation of the interfacial fluid in terms of an enhanced viscosity of the confined film.[44, 45] To test this hypothesis against our data, we start from a model derived by Chan and Horn,[46, 47] which accounts for the viscous force between approaching crossed-cylinder surfaces (radius *R*) as a function of their separation and the fluid viscosity, $\eta$:

$$\text{For } D > b: \qquad \mathrm{F} = -\frac{6\pi\eta R^2}{D-b}\frac{dD}{dt} \qquad \text{Eq. 3}$$

To allow for the hypothecated interfacial network, we extend Eq.3 to incorporate a confinement-dependent viscosity; $\eta = \eta(D)$:

$$\eta(D) = \eta_{bulk} + He^{-(D-b)/\lambda} \qquad \text{Eq. 4}$$

where $\eta(D)$ and $\eta_{bulk}$ are viscosities of the confined film and bulk fluid respectively, and $H$ is the limiting excess viscosity at the slip plane which is located at a distance *b* from each surface. The enhancement of viscosity extends over a characteristic length $\lambda$. This is the simplest model found to provide a satisfactory empirical fit to the data, including the observed velocity dependence, in the range where viscous (hysteretic) forces were observed. Equation 3 may be solved by numerical integration to generate a fit for the measured interactions. Note that the vdW interaction is ignored for simplicity as it has negligible impact on the total interaction at large separations.

A representative fit is shown in Figure 2A, illustrating a good fit to the measured data down to $D \sim$ 10nm (solid line). Also plotted in 3A is the force calculated for a film retaining its bulk viscosity at all separations, $\eta(D) = \eta_{bulk}$ (dashed line), illustrating the significant effect of enhanced viscosity in confinement. A key prediction of our model is that the measured viscous force is velocity dependent ($\mathrm{F} \propto dD/dt$). Figure 2B shows representative force profiles acquired at different applied speed. Each dataset was fitted individually to extract the excess viscosity, slip length, and decay length. The resulting parameters were averaged and then fixed across all datasets. Using this single parameter set, we calculated the viscous force according to Equations 3 and 4 and compared it with the measured force (solid lines in 3B). The model can clearly capture both the magnitude and the expected speed dependence of the viscous force well for *D* > 10nm, supporting the viscous force interpretation.

For the thinnest films, with *D* < 10 nm, we observe a different scaling of force with confinement and the repulsive force becomes reversible (*i.e.* 'in' and 'out' profiles coincide), indicating elastic rather than viscous mechanical properties in this range. This is illustrated for one example run in Figure 2C, where the viscous region (10 < *D* < 50 nm) displays in/out hysteresis while the thinnest films (*D* < 10 nm) are reversible.  This indicates that dissipation due to viscous flow is no longer dominant; instead, the force is consistent with an elastic response arising from deformation of a confined interfacial layer on the timescale of our experiment ($10^2$ s). If hydrated networks are stabilised at the solid-liquid interface, we estimate that the thickness of this elastic, film-like structure is on the order of $\sim$5 nm on each surface; this value is in quantitative agreement with recent scattering experiments of ion clusters observed in the bulk solution.[10].

Within the above model, the excess viscosity of the interfacial networks is 7.7 ± 2.7 mPas for the 0.28 M $CaCl_2$ solution with a characteristic length of 34.3 ± 8.7 nm. The effective viscosity is roughly one order of magnitude higher than the experimental bulk viscosity (1.06 mPas) and significantly smaller than the viscosity of supersaturated dense liquid droplets estimated by liquid-phase transmission electron microscopy, but comparable with recent atomic force microscopy measurements.[5, 45] To confirm our hypothesis of calcium carbonate hydrated networks are present, we have further studied the effects of pH and competing cations as discussed below.

**Chemical speciation alters viscoelastic behaviour**

Our recent work has highlighted the critical role of pH in governing interfacial speciation of carbonate ions.[22] If the viscous force we observed is due to calcium carbonate networks, then it should be sensitive to pH.[8, 9] We have performed measurements at lower and higher pH by adding trace HCl and KOH (0.001 M) to the 0.28 M $CaCl_2$ solution. Representative interaction profiles are shown in Figure 3. While pH clearly regulates the interaction between mica surfaces, a long-range repulsion is always present. The magnitude of measured interaction at short distance is difficult to disentangle due to confounding variables such as surface potential, speciation, and local concentrations,[22] which are all uncontrolled. Nevertheless, the long-range repulsion can be interpreted as a viscous force by fitting Equations 3 and 4. We found that at lower pH (= 3.2), the excess viscosity and its decay length were found to be highly variable (13.7 ± 10.4 mPa s and 37.0 ± 28.2 nm). This fluctuation likely arises from the shift in the bicarbonate/carbonate equilibrium as pH is lowered, resulting in the formation of a patchier network that is more sensitive to the local surface and solution conditions.[8, 48] Conversely, higher pH should favour the formation of more stable interfacial calcium carbonate hydrated networks by increasing the concentration of carbonate. However, by increasing the pH to 9.2, the solution became super saturated, and interaction profiles can no longer be measured reproducibly (see SI). This led to a much wider distribution of measured interaction profiles, indicating that nucleation of amorphous calcium mineral is likely to have occurred.[37]

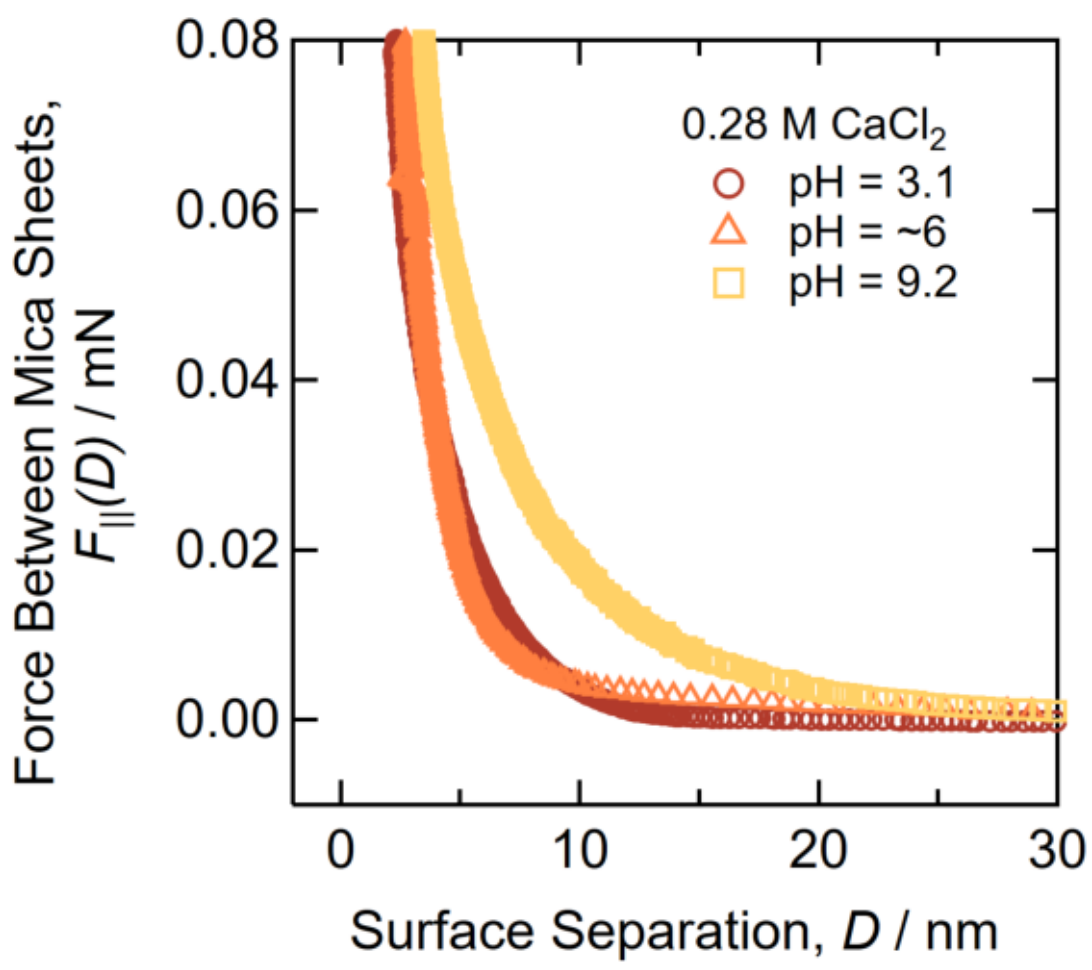

**Figure 3.** pH dependence of the non-DLVO repulsive interaction measured between mica sheets across calcium chloride solutions. The decay length clearly differs, but this long-range feature is always present. pH value of the neutral (unadjusted) solution varied between 6.2 and 5.6 throughout the measurement period.

**Competing cations can displace the interfacial networks**

The mica clay interfaces used in our measurements is a layered aluminosilicate mineral with interlayer potassium ions balancing the net negative charge of the silicate sheets. The negative charge arises from dissociation of potassium ions from the surface layer once in contact with electrolyte solutions.[42] Since potassium is the native ion for muscovite mica, we hypothesised that it has a strong affinity for the silicate surface and may act as a competing ion to displace the interfacial calcium carbonate networks. We have studied a range $CaCl_2$ solutions with added KCl ranging from equimolar to a 1:3 mole ratio (*i.e*. at equal ionic strength). A representative interaction profile of 0.28 M $CaCl_2$ + 0.9 M KCl is shown in Figure 4A.

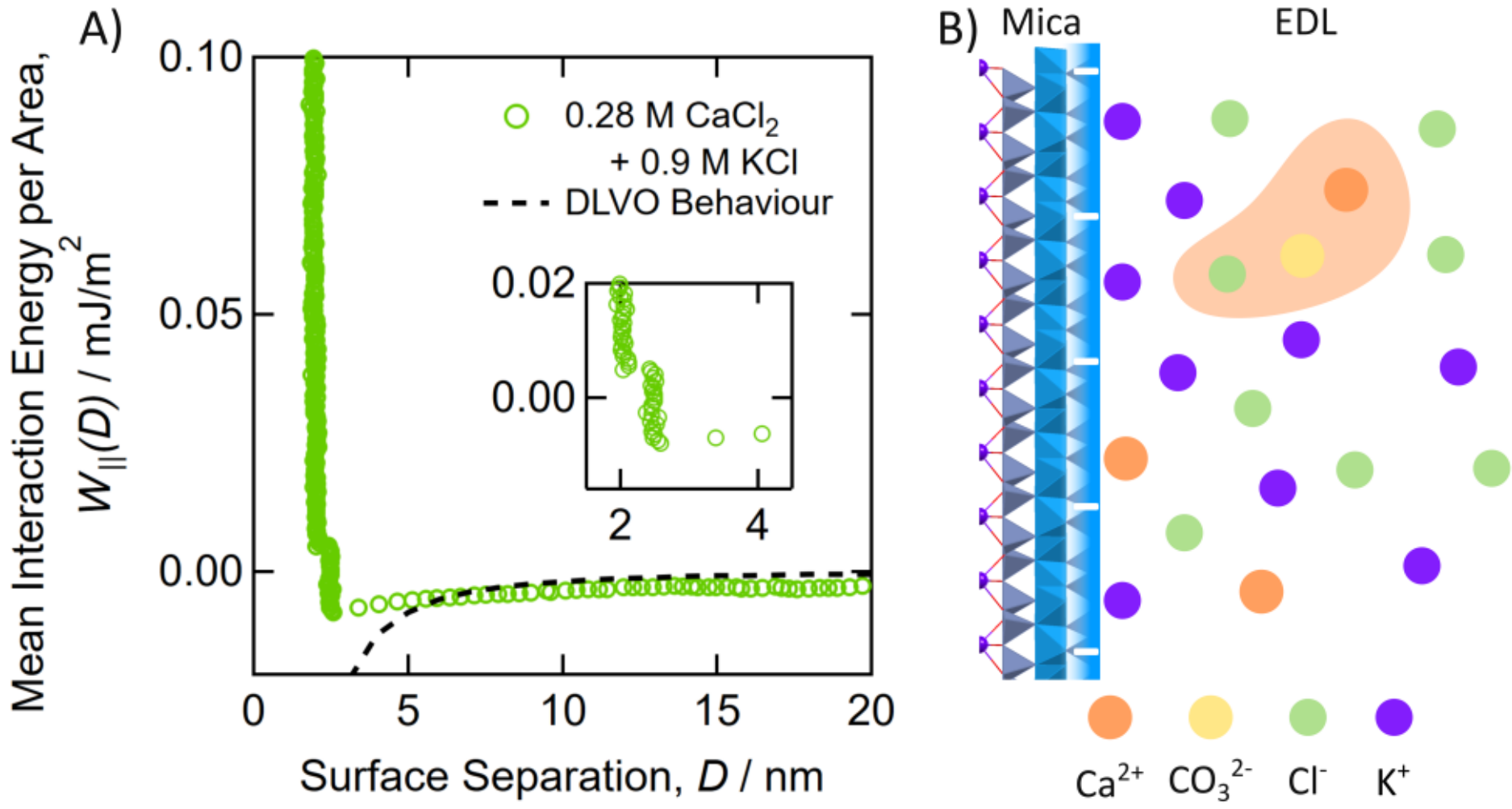

**Figure 4.** A) Interaction between mica sheets across a solution containing 0.28 M calcium chloride and 0.9 M potassium chloride. The non-DLVO long-range repulsive force is displaced by classical behaviours. A vdW jump-in of > 2 nm is observed, followed by a molecular step of around 0.5 nm. This step-size is consistent with the dimensions of solvated potassium and consistent with literature. B) Schematic of how potassium ions can disrupt and displace the hydrated calcium network.

At high KCl concentration, the long-range viscous force disappears. The interaction profile between mica surfaces across this $CaCl_2$ + KCl solution appears restored to classical behaviours, similar to the $MgCl_2$ solution in Figure 1B. While the long-range viscous force is disrupted, a small step of 0.5 nm is observed near contact after the vdW jump-in. This can be attributed to a layer of weakly adsorbed hydrated potassium ions, which was previously reported at high KCl concentration (2.5 M).[23] This confirms that the interfacial calcium ions are displaced by potassium, and consequently the long-range viscous force is no longer measurable.

**Conclusion and outlook**

The nucleation pathway of calcium minerals from bulk solutions remains a subject of debate.[5-12, 16] In particular, distinguishing between classical nucleation pathways and models invoking prenucleation clusters requires determining whether stable ion assemblies can exist within the undersaturated regime. We studied calcium-containing solutions in equilibrium with air to investigate undersaturated solution behaviours at the mineral-solution interface and under controlled confinement. We have experimentally measured the interaction potential between negatively charged mica surfaces across calcium chloride and nitrate solutions at various concentrations, pH, and in the presence of potassium chloride. These measurements reveal the emergence of a long-range repulsive force specific to calcium-containing solutions beyond 0.1 M.

We interpret this force as arising from stable hydrated ion networks at the mineral–solution interface and characterise their viscoelastic behaviour. Similar structures have recently been reported in atomic force microscopy studies on trivalent ions at mica surfaces,[44] as well as at boehmite-water interfaces,[45] supporting our interpretation. The effective viscosity of these interfacial networks is approximately one order of magnitude greater than that of the bulk solution. The pH study further demonstrated that this interfacial feature persists across the investigated pH window (3.1-9.2), despite substantial changes in carbonate speciation. However, the molecular mechanism of this pH-sensitivity is difficult to disentangle due to confounding surface and solution properties that are also sensitive to pH. By introducing potassium ions, the native counterions of mica surfaces, we demonstrated that the interfacial networks can be disrupted. This suggests that, in addition to carbonate speciation, the stability of these viscoelastic structures is governed by competitive adsorption at the mineral surface.

Our findings provide experimental evidence that stable ion networks can emerge in calcium-containing solutions at mineral interfaces under conditions that remain undersaturated with respect to bulk $CaCO_3$ precipitation. These observations support the possibility that intermediate ion assemblies may form prior to nucleation and that interfaces and confinement can stabilise such structures beyond what is expected from bulk thermodynamics alone. While the present study establishes the existence and viscoelastic nature of these hydrated calcium networks, further investigation of their shear rheology, water dynamics, temperature dependence, and structural organisation will be necessary to determine their relationship to prenucleation clusters, amorphous intermediates, and nonclassical crystallisation pathways more broadly. More generally, our results highlight the central role of interfaces and confinement in directing mineralisation processes and may ultimately improve our ability to predict and control crystallisation in confined environments relevant to biomineralisation, geochemistry, and scalable carbon storage.

## Methods

We used a surface force balance (SFB) to measure the interaction between two identical ruby muscovite mica (S&J Trading Inc.) surfaces against their separation. Spatial resolution of approximately 0.1 nm was achieved by using white light interferometry and a digital camera (QImaging Retiga R6 charged-coupled device, frame rate ∼5 Hz). The experimental setup is shown in Figure 1A, and has been described in detail in a recent review.[25] Electrolyte solutions were prepared in a dust-free environment with all glassware cleaned in piranha solution (a 3:1 mixture of sulfuric acid (supplier, 95 %) and hydrogen peroxide (Sigma Aldrich, 30 %)) prior to use. Prior to injection into the SFB, the electrolyte solutions were filtered using 100 nm nylon syringe filter (Cytiva, Whatman™ Puradisc 13). Around 0.3 mL of the filtered electrolyte was injected between the mica surfaces mounted inside the SFB. Calcium chloride hydrate (99.9965%, Thermo Scientific), magnesium chloride hexahydrate (99.999%, Thermo Scientific), calcium nitrate tetrahydrate (99.98%, Thermo Scientific), hydrochloric acid (∼37%, Analytical grade, Fisher Chemical), potassium hydroxide (pellet, >84%, Merck), and potassium chloride (99.997%, Alfra Aesar) were all used as received. Ultrapure water was produced by Milli-Q® IQ 7003 with resistivity of 18.2 MΩ cm and total organic carbon less than 5 parts per billion. To ensure that the investigated solutions were in equilibrium with atmospheric $CO_2$, solutions were prepared c. 30 minutes in advance of the measurement and allowed to equilibrate. During the measurements, the solutions were always in contact with atmosphere. Water content of the calcium chloride hydrate was determined by thermogravimetric analysis (Mettler Toledo TGA/DSC 1 System), pH was measured by a HI5221 pH meter (HANNA® Instruments), and refractive index was measured by an Abbe 60 refractometer (Bellingham + Stanley).

## Author Contributions

S.M. and S.P. conceived the research idea. S.M. and K.J.A. performed the experiments. S.M., T. S. G. and S.P. analysed the data. S.M. and S.P. secured funding and resources. S.M. wrote the original manuscript and all authors contributed to review and editing.

**Conflict of Interest:**

There are no conflicts of interest.

**Data availability**

All data supporting this article have been included as part of the Supplementary Information available at: xxxxx. Full dataset supporting the findings of this study are available from the Oxford University Research Archive: xxxxx.

**Acknowledgement**

The authors gratefully acknowledge funding from the European Research Council under grant 101001346 ELECTROLYTE. S. M. is supported by a Career Development Research Fellowship from St. John's College, Oxford. K. J. A. would like to acknowledge support from The Oxford-The Queen's College Graduate Scholarship in partnership with the Clarendon Fund, University of Oxford.

**Abbreviations**

CNT, Classical Nucleation Theory; SFB, Surface Force Balance; vdW, van der Waals; EDL, Electrical Double Layer; DLVO, Derjaguin–Landau–Verwey–Overbeek.